\documentclass
[prb,aps,twocolumn,superscriptaddress,showpacs,amsfonts,amsmath,amssymb,showkeys,floatfix]{revtex4}
\usepackage{bm}
\usepackage{pstricks}
\usepackage{setspace}
\usepackage{graphicx}
\begin{document}
\title{Monte Carlo study of the two--dimensional site--diluted dipolar Ising model}
\date{\today}
\author{Juan J. Alonso}
\email[E-mail address: ] {jjalonso@uma.es}
\affiliation{F\'{\i}sica Aplicada I, Universidad de M\'alaga, 29071 M\'alaga, Spain}
\author{B. All\'es}
\email[E-mail address: ] {alles@df.unipi.it}
\affiliation{INFN--Sezione di Pisa, Largo Pontecorvo 3, 56127 Pisa, Italy}

\date{\today}

\begin{abstract}
By tempered Monte Carlo simulations, we study 2D site--diluted dipolar Ising systems.
Dipoles are randomly placed on a fraction $x$ of all $L^2$ sites in a square lattice, and
point along a common crystalline axis.  For $x_c< x\le1$, where $x_c = 0.79(5)$, we find an 
antiferromagnetic phase below a temperature which vanishes as $x  \to x_c$ from above. 
At lower values of $x$, we study {\it (i)} distributions of the spin--glass (SG) overlap $q$, {\it (ii)} their relative mean square 
deviation $\Delta_q^2$ and kurtosis and {\it (iii)} $\xi_L/L$, where $\xi_L$ is a
SG correlation length. From their variation with temperature and system size, we find that
the paramagnetic phase covers the entire $T>0$ range. Our results enable us to obtain an estimate of
the critical exponent associated to the correlation length at $T=0$, $1/\nu=0.35(10)$.
\end{abstract}

\maketitle

\section{Introduction}
\label{intro}

In the last years, there has been a renewed interest in systems of interacting dipoles (SIDs).
This is in part due to recent advances in nanoscience\cite{nanoscience} which make
realizations of assemblies of magnetic nanoparticles available.\cite{np,sachan} Empirically, these systems
show a rich collective behavior in which the dipole--dipole interaction plays a key role that
can be observed at low (but experimentally accessible) temperatures.  Dipoles forming crystalline arrays
exhibit long--range ferro or antiferromagnetic order that depends crucially on lattice geometry\cite{lutt,odip}
because of geometric frustration caused by the spatial variations of the directions of dipolar fields.
Two--dimensional (2D) arrays of cobalt--ferrite  and Co nanoparticles
placed on hexagonal arrays have been found to exhibit in--plane short--range ferromagnetic order.\cite{dos}
On the contrary, arrays on a square lattice composed of Mn$\,$As ferromagnetic nanodisks epitaxially
grown on a substrate  exhibit collinear AF patterns.\cite{dosAF}

Magnetic ordering of SIDs depends also on anisotropy.  On the one hand, dipolar--dipolar
interactions create effective anisotropies that in square lattices, for example,
push spins to lie on the plane of the lattice.\cite{debell} On the other hand, magnetocrystalline
site--anisotropy energies of the crystallites that form the nanoparticles are often
greater than dipolar--dipolar interparticle energies.  This is the case of the arrays of Mn$\,$As ferromagnetic 
nanodisks we mention above, that  behave as a system of Ising dipoles
with their magnetic moment rigidly aligned along the in--plane crystalline easy axes of the nanodisks.\cite{dosAF}
In such a case, the resulting magnetic order depends on the competition of dipolar and
anisotropic energies. Crystalline Ising  dipolar systems (IDSs) are reasonable models for these
planar systems.\cite{jensen} Some ferroelectrics,\cite{ferroel} insulating magnetic salts as
Li$\,$Ho$\,$F$_4$,  as well as some three--dimensional (3D) crystals of organometallic
molecules \cite{nanomag} are known to be well described by arrays of IDSs.\cite{jensen,bel}

SIDs in disordered spatial arrangements are  particularly interesting. The presence of spatial
disorder, together with the geometric frustration generated by dipolar interactions, gives rise
to random frustration that may result in SG behavior. In fact, some non--equilibrium SG behavior
(like time dependent susceptibilities and memory effects) has been observed in experiments with systems
of randomly placed nanoparticles or very diluted magnetic crystals.\cite{ferroel,experiments}
Furthermore, Monte Carlo (MC) simulations have given clear evidence of the existence of a transition at finite temperature 
$T_{SG}$ from a paramagnetic to an equilibrium SG phase in systems of randomly oriented 
axis dipoles (RADs) placed either on fully occupied or on
diluted simple cubic (SC) 3D lattices, and $T_{SG}=0$ instead for 2D square lattices.\cite{rad} Recent numerical work
has reported a SG transition in a model of parallel axis dipoles (PADs) placed on a lattice that approximates
that of the diluted\cite{gin} Li$\,$Ho$_x$Y$_{1-x}\,$F$_4$, a material  for which such a transition 
has been reported,\cite{wu} (albeit not without some controversy\cite{barbara}). By MC simulation
the whole phase diagram of site--diluted PADs placed on a 3D SC lattice has been obtained\cite{2009b} as a
function of the concentration $x$. It includes a SG phase for $0<x\lesssim0.65$ which,
strikingly, has been found to behave marginally, that is, it has quasi--long range 
order, as in the 2D XY model.\cite{xy} This is contrary to theoretical expectations,\cite{bray, katz} 
that SG systems with long--range interactions may behave as short--range Edwards--Anderson 
(EA) models,\cite{3dEA} which in 3D are believed to have a  SG phase with a non--vanishing order 
parameter (according to the RSB\cite{RSB} or droplet \cite{droplet} pictures of SGs). 

Then,  in order to get a deeper understanding of SG systems beyond the already extensively
studied random--bond models with short--range interactions, it makes sense to analyze the behavior of the 2D PAD
model and compare it with the short--range 2D EA model.
The latter had been found to have an algebraic divergence at $T_{SG}=0$
with critical exponent\cite{2dEAold} $1/\nu=0.50(5)$, although more recent simulations for larger systems and lower
temperatures give a value of $1/\nu=0.29(4)$ for Gaussian interactions.\cite{2dEAnew}

Our purpose is the study by MC simulations the phase diagram of a site--diluted system of magnetic dipoles.
They are placed at random on the sites of a square lattice and point up or down along a given principal axis.
Since in the limit of low  concentrations every detail of the lattice is expected to become
irrelevant,\cite{2009b} our results
have direct connection with some of the work we describe above. Our intention is to search
for the temperature $T_{SG}$ of a possible SG transition and study the related divergence of the correlation
length. Further, we aim to study whether the diluted PAD model belongs to the same universality
class recently conjectured,\cite{jorg} though not reliably shown by MC simulations,\cite{bis} for the 
set of 2D EA Ising models with varying quenched disorder.

The plan of the paper is as follows. In Section~\ref{mm} we define the model and give details on the parallel tempered
Monte Carlo (TMC) algorithm\cite{tempered} used for updating.
We also define the quantities we calculate. They include the spin overlap\cite{ea} $q$
and a  correlation length\cite{longi,3dEA} $\xi_L$.
In Section~\ref{resultsA} we give results for the dipolar AF phase for $x>x_c$, where $x_c= 0.79(5)$, as well as for
its nature and boundary. In Section~\ref{resultsB}, numerical results are shown for distributions of $q$ and $\xi_L/L$
at $x=0.2$ and $0.5$. We examine the evidence against the existence of a finite temperature SG phase transition
when $x<x_c$: {\it (i)} the mean values $\langle \mid q\mid\rangle$ and $\langle q^2\rangle$ decrease faster
than algebraically with $L$ as $L$ increases for $T/x\gtrsim0.3$, {\it (ii)} double peaked, but wide, distributions
of $q/\langle \mid q\mid\rangle $ change with $L$ for temperatures as low as $T/x=0.4$, 
and {\it (iii)} kurtosis and $\xi_L/L$  decrease with $L$ at all $T$ and do not cross, 
as it would be expected for a finite temperature transition. Scaling plots for $g$ and  $\xi_L/L$ are given
in Section~\ref{resultsC}. Our results are consistent with a ratio $\xi_L/L$ that diverges with
exponent $1/\nu=0.35(10)$. Results are summarized in Section~\ref{conclusions}.

\section{model, method, and measured quantities}
\label{mm}
\subsection{Model}
We treat site--diluted systems of Ising magnetic dipoles (also named spins in this paper) on a 2D square lattice.
At each lattice site a dipole is placed with probability $x$. Then, the number $N$ of spins on
the lattice is less than $L^2$ ($L$ is the lateral size of the lattice)
approximately by a factor $x$. Site $i$ is said {\it occupied} if it
contains one spin. All dipoles are parallel and point along the $Y$ axis of the lattice. This axis
shall be called {\it spin axis}. The Hamiltonian is given by,
\begin{equation}
{H}=\frac{1}{2}\sum_{ i\not=j}
T_{ij}\sigma_i \sigma_j\,,
\label{hamiltonian}
\end{equation}
where the sum runs over all occupied sites $i$ and $j$ except $i=j$,
$\sigma_i=\pm 1$ on any occupied site $i$ and
\begin{equation}
T_{ij}=\varepsilon_a
(a/r_{ij})^3(1-3y_{ij}^2/r_{ij}^2).
\label{T}
\end{equation}
If ${\bf r}_{ij}$ is the vector joining sites $i$ and $j$, then $r_{ij}=\Vert{\bf r}_{ij}\Vert$ is its
modulus and $y_{ij}$ its $Y$ component. $\varepsilon_a$ is an energy and $a$ is the lattice spacing.
In the following all temperatures and energies shall be given respectively in units of
$\varepsilon_a/k_B$ ($k_B$ is the Boltzmann constant) and $\varepsilon_a$.

Due to the long--range nature of the dipolar interactions, we are able to simulate on rather small lattice
sizes ($L\leq 32$).

Strength $T_{ij}$ is the usual long--range dipole--dipole interaction. Note that
$T_{ij}$ signs are not distributed at random, but depend on the
orientation of ${\bf r}_{ij}$ vectors on the lattice. Randomness in our model
arises only through the introduction of the probability $x$ for placing dipoles.
This is to be contrasted with random--bond EA Ising models with bond strengths
$J_{ij}=\varepsilon_{ij}/r_{ij}^\mu$ where $\varepsilon_{ij}$ are chosen at random from
a bimodal or Gaussian distribution with zero mean\cite{katz} and $\mu$ is a real exponent. This is why PADs exhibit AF order
at high concentration in contrast to these models that do not. Similar statements apply when our PAD model is
compared with a random--axes dipolar model (RAD), in which Ising
dipoles lie along directions chosen at random for each site.\cite{rad}

\subsection{Method}
 
Periodic boundary conditions (PBC) are imposed. Spins on occupied sites $i$ have been allowed to interact only with
spins $j$ within an $L\times L$ squared box centered on site $i$. This method unambiguously defines the vector ${\bf r}_{ij}$
to be used in~(\ref{T}) and also excludes interactions with spins belonging to the repeated
copies of the lattice that appear beyond the boundary. Because of the
long--range nature of dipolar interactions, contributions from beyond this box would have been
taken into account (for example by means of Ewald's summations\cite{ewald})
if spins were to form ferromagnetic domains. They do not do so in our PAD model. In all
simulations presented in this work we have found  $T\chi_{F} \lesssim 1$, where $\chi_{F}$ is the
ferromagnetic susceptibility. Therefore those contributions do not affect the thermodynamic limit
regardless of the kind of phase the system lies (paramagnetic, AF or SG).
Some details on this point are found in.\cite{2009b}

\begin{table}\footnotesize
\caption{Simulation parameters. $x$ is the probability for sites to be occupied
with a magnetic dipole; $L$ is the lateral lattice size; $\Delta T$ is the temperature step
in the TMC runs; $T_1$ and $T_n$ are the highest and lowest temperatures, respectively;
$N_r$ is the number  of pairs of quenched disordered samples; $t_0$ is the
number of MC sweeps. The measuring time interval is $[t_0, 2t_0]$ in all cases.
}
\begin{tabular}{p{1.2cm} p{1.0cm} p{1.0cm} p{1.0cm} p{1.0cm} p{1.0cm} p{1.0cm}}
\hline\hline
\multicolumn{7}{c}
{$x=0.2$, $\Delta T=0.02$, $T_1=0.6 $, $t_{0}=4 \times  10^7$}\\
\hline
$L$    &8     &12   &16    &20& 24&32 \\
$T_n$&0.04&0.04&0.04&0.04&0.04& 0.08\\
$N_r$&2400&550&1500& 650&700& 200\\
\hline 
\multicolumn{7}{c}
{$x=0.5$, $\Delta T=0.05$, $T_1=2$, $T_{n}=0.1$, $t_{0}=8 \times  10^6$}\\
\hline
$L$       & 8        &16      & 20    & 24   \\
$N_r$   &2500   &2500  &350  &250\\
\hline 
\multicolumn{7}{c}
{$x=0.6$, $T_{n}=0.2$, $t_{0}=4 \times  10^6$}\\
\hline
$L$          		& 8     &16      & 20    & 24 \\
$T_1$  		&3      &3        &2       & 2\\
$\Delta T$         &0.2   &0.2     &0.1    &0.1\\
$N_r$   		&1200 &300   &300   &300\\
\hline 
\multicolumn{7}{c}
{$x=0.7$, $\Delta T=0.1$, $T_1=2$, $T_{n}=0.2$}\\
\hline
$L$          		& 8     &16      & 20     & 24 \\
$t_{0}$              &$8 \times  10^5$&$8 \times  10^5$&$4 \times  10^6$&$4 \times  10^6$\\
$N_r$   		&4200 &2200    &400    &100\\
\hline 
\multicolumn{7}{c}
{$x=0.8$, $\Delta T=0.1$, $T_1=3$,  $T_{n}=0.2$}\\
\hline
$L$          & 8        &16      & 20    & 24   \\
$t_{0}$              &$8 \times  10^5$&$8 \times  10^5$&$4 \times  10^6$&$4 \times  10^6$\\
$N_r$   &4500   &1200  &500  &350\\
\hline 
\multicolumn{7}{c}
{$x=0.86$, $\Delta T=0.1$, $T_1=3$, $T_{n}=0.2$, $t_{0}=8 \times  10^5$}\\
\hline
$L$          & 8        &16      & 20    &24   \\
$N_r$   &3000  &400    &300   &70\\
\hline 
\multicolumn{7}{c}
{$x=0.9$, $\Delta T=0.1$, $T_1=3$, $T_{n}=0.2$, $t_{0}=8 \times  10^5$}\\
\hline
$L$           & 8       &16      & 20    &24   \\
$N_r$   &2000  &250    &250   &800\\
\hline\hline 
\end{tabular}
\label{table}
\end{table}

In order to circumvent large energy barriers that could slow down the evolution of the system, in particular
from certain states representing minima of the energy (mainly at low temperatures), we have used the TMC
algorithm.\cite{tempered} It consists in running in parallel a set of $n$ identical systems at
equally spaced temperatures $T_i$, given by $T_i=T_1- (i-1) \Delta T$ ($i=1, \cdots, n$
and $\Delta T>0$) where each system $i$ is cyclically allowed to exchange its state with system $i+1$.
Each system evolves independently by use of the standard single--spin--flip Metropolis
algorithm\cite{mc} and whenever a single flip is accepted, all dipolar fields throughout the entire
lattice are updated.

In detail the procedure is as follows:~\cite{2009b,rad}
{\it (1)} a cycle on $i$ is run from $i=1$ to $i=n$;
{\it (2)} when the cycle arrives at system $i$, 8 Metropolis steps are applied on it;
{\it (3)} next, a chance is given to systems $i$ and $i+1$ to exchange their configurations
(note that at this moment system $i+1$ has undergone 8 Metropolis steps less than system $i$).
The exchange is accepted with probability $P_{TMC}=1$ if $\delta E= E_{i}-E_{i+1}<0$ or
$P_{TMC}=\exp (-\Delta \beta \delta E)$ otherwise. Here $E_i$ is the numerical value
of Hamiltonian~(\ref{hamiltonian}) for system $i$ and $\Delta \beta=1/T_{i+1}-1/T_{i}$;
{\it (4)} 8 Metropolis steps are applied on system $i+1$ (regardless of the fact that the
previous exchange have or have not been performed);
{\it (5)} the above exchange is tried between systems $i+1$ and $i+2$;
{\it (6)} the cycle ends after the 8 Monte Carlo steps for $i=n$, after which no configuration exchange is tried.

Since\cite{2009b} in 3D $T_{SG}\sim x$ and the purpose of TMC is to overcome energy barriers that
could be as high as $T_{SG}$, then we found necessary to choose the highest temperature $T_{1}\gtrsim 2 x$.
It is also important to take $\Delta T$ small enough to allow frequent state exchanges between systems.
This is fulfilled by taking $\Delta T \lesssim T/\sqrt{c_s N}$ where $c_s$ is the specific heat per spin.
We choose appropriate values for $\Delta T$ from inspection of plots (not shown) of the specific heat vs
$T$ in preliminary simulations of the smaller systems.\cite{2009b}

Initially the $n$ configurations were completely disordered. For details on how we chose equilibration
times $t_0$ see Section \ref{meas}. Time $t_0$ is particularly large outside the AF region, varying
from at least $4 \times 10^6$ MC sweeps for $x=0.7$ and a number of dipoles $N\ge300$ up to $4 \times 10^7$
sweeps for $x=0.2$ and $N=200$. Instead, $t_0$ in the AF zone is as low as $8 \times 10^5$ for $x\ge 0.86$.
Thermal averages were calculated over the time range $[t_0,2 t_0]$.  We further
averaged over $N_r$ samples with different realizations of disorder. Each realization was run twice
to permit the calculation of overlapping parameters (see  Section \ref{meas}).
Values of the parameters for all TMC runs are given in Table~I.

\subsection{ Measured quantities}
\label{meas}
Measurements were performed after two averagings: first over thermalized configurations and secondly over different realizations
of the quenched disorder.

We begin by presenting the specific heat. It was extracted from the slope of the energy as a function of the temperature.

As for the staggered magnetization, also for a PAD model on the square lattice we find appropriate to define it as\cite{odip}
\begin{equation}
m=N^{-1} \sum_i (-1)^{x_i} \sigma_i\,,
\label{phi}
\end{equation}
where $x_i$ is the $X$ coordinate of site $i$. We calculated the probability distribution $P(m)$, as well as
the moments $m_{1}=\langle|m|\rangle$, $m_{2}=\langle m^{2} \rangle$, and  $m_{4}=\langle m^{4} \rangle$,
where $\langle\dots\rangle$ stands for the above--defined double averages.
From these moments we calculated the kurtosis (known also as Binder's cumulant)
of $P(m)$ as $g_{m}=(3-m_{4}/m_{2}^2)/2$. All these quantities have proven to be good
signatures for possible AF--paramagnetic phase transitions.

\begin{figure}[!t]
\includegraphics*[width=80mm]{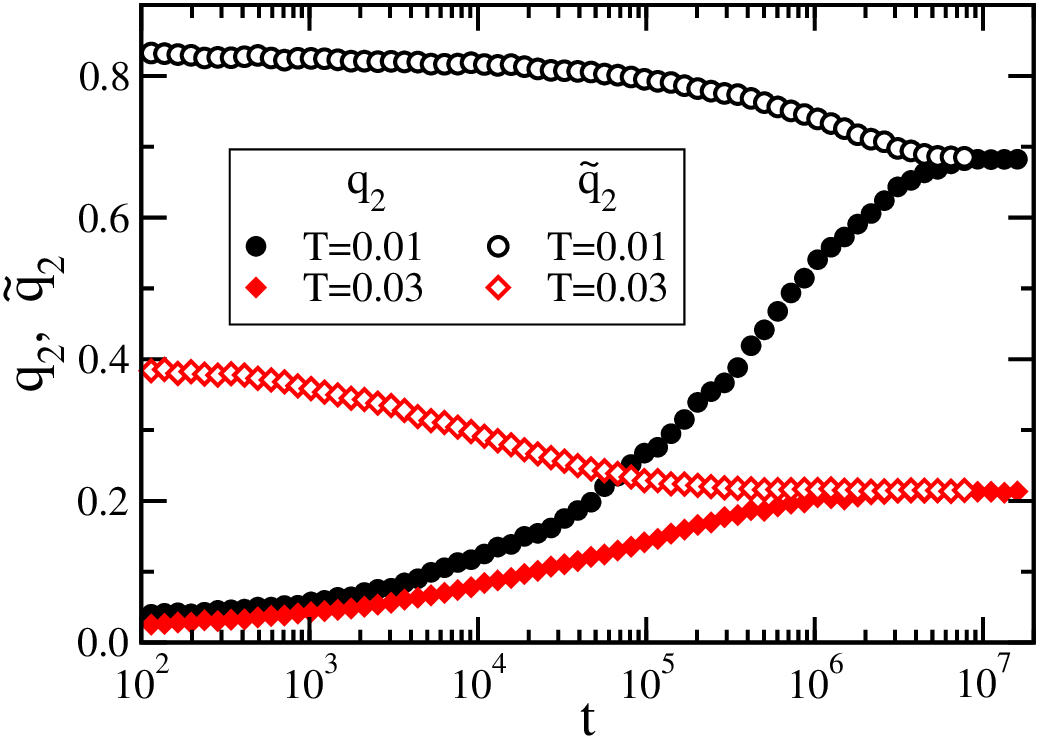}
\caption{(Color online)
Semilog plots of $\widetilde q_2(t_0,t)$ and $q_2$ vs time $t$ (in MC sweeps) for systems of $24 \times 24$ sites
and concentration $x=0.5$  at the values of $T$ shown in the legend. Here, $q_2$ comes from averages
over time, starting from an initial random spin configuration at $t=0$.
Here $t_0=8\times10^6$ MC sweeps. Data points at time $t$  from an average over the time interval $[t,1.2t]$ and
over $500$ system samples.
}
\label{time}
\end{figure}

In order to look for SG behavior, we also calculated
the Edwards--Anderson overlap parameter between two independent equilibrium
configurations obtained from a pair of identical replicas evolving
independently in time,\cite{ea}
\begin{equation} 
q=N^{-1} \sum_j \phi_j\,,
\label{q1}
\end{equation}
where 
\begin{equation} 
\phi_j=\sigma^{(1)}_j\sigma^{(2)}_j\,,
\label{phil}
\end{equation}
$\sigma^{(1)}_j$ and $\sigma^{(2)}_j$ being the spins on site $j$ of
replicas labelled as $(1)$ and $(2)$. Clearly, $q$ is a measure
of the spin configuration overlap between the two replicas. As done for $m$, we also calculated
the probability distribution $P(q)$ as well as the moments $q_1=\langle |q|\rangle $,
$q_2=\langle q^2\rangle $, and $q_4=\langle q^4\rangle$. The SG susceptibility $\chi_{SG}$
is given by $Nq_2$. Finally, we also make use of the relative mean square deviation of $q$,
$\Delta^2_q= q_2/q_1^2-1$, and kurtosis $g=(3-q_{4}/q_{2}^2)/2$.

Let us explain now how $t_0$ was extracted.
To make sure that equilibrium was reached, plots of $q_2$ and energy (average of $H$) were made
over time intervals $[t,1.2 t]$, not starting at $t=t_0$, as we do everywhere else, but starting
at $t=0$, from an initial random spin configuration. Semilog plots of $q_2$ versus $t$
displayed in Fig.~\ref{time} for $x=0.5$, $L=24$ and low temperatures show that a stationary state
is reached only after some millions of MC sweeps. In order to check whether this state is truly
in equilibrium, we define a time dependent spin overlap $\widetilde{q}$, not among pairs of
identical systems, but among spin configurations of the same system at two different
times $t_0$ and $t_1=t_0+ t$ of the same TMC run,
\begin{equation}
\widetilde q(t_0,t)= N^{-1} \sum_j \sigma_j(t_0)\sigma_j( t_0+t).
\label{phi0}
\end{equation}

Let $\widetilde{q}_2(t_0,t)=\langle\left( \widetilde q(t_0,t) \right)^2 \rangle$. Suppose thermal equilibrium
is reached at a time $t'$. Then, $\widetilde{q}_2(t_0,t)\rightarrow q_2$ as $t \to t'$ provided that $t_{0}\gtrsim t'$.
Plots of $\widetilde{q}_2(t_0,t)$ vs $t$, for $10^{-5}t_0<t<t_0$ and $t_0=8 \times 10^6$ MC sweeps, are shown in
Fig.~\ref{time}. Note that both quantities, $q_{2}$ and  $\widetilde{q}_{2}$ become approximately equal
when $t\gtrsim 10^6$ MC sweeps. In order to obtain equilibrium results, we have always chosen
sufficiently large values of $t_0$  to make sure that, within errors,   $\widetilde{q}_2(t_0,t) = q_2$ for $t\gtrsim t_{0}$.
All values of $t_0$ are given in Table~I.

\begin{figure}[!t]
\includegraphics*[width=80mm]{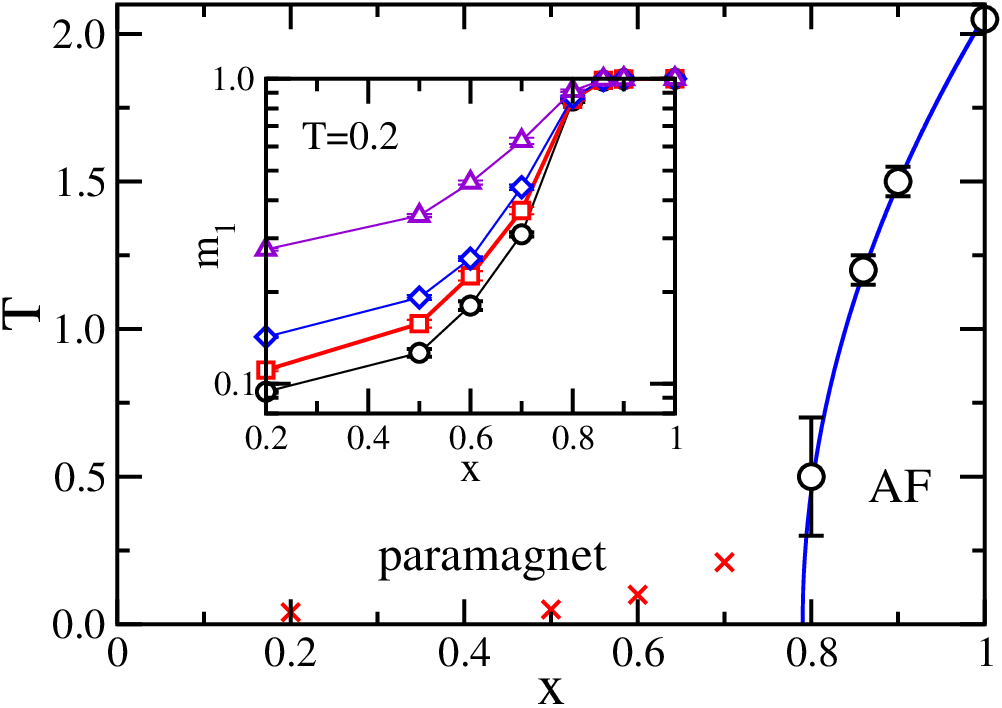}
\caption{(Color online) Phase diagram of the 2D PAD model.
$\circ$ stand for the N\'eel temperature $T_{N}$ and $\times$ for temperatures below which we
cannot completely rule out a SG phase (see Section~\ref{results}). The full line for the phase boundary
between the paramagnetic and AF phases is a fit to the data points given by
$T_{AF} \simeq 4.5 (x-x_c)^{1/2}$, where $x_c=0.79$. In the inset, $m_1$ versus $x$ for $T=0.2$.
$\circ$, $\square$, $\diamond$, and $\triangle$,  stand for $L=24, 20, 16$, and $8$ respectively.}
\label{fases}
\end{figure}

In addition, we calculated a so--called correlation length for finite systems,
\begin{equation} 
\xi_{x,L}=\frac {1 } {2 \sin(k /2)}  { \left[ \frac{q_2} { \langle\mid q({\bf k})  \mid ^2   \rangle}  -1 \right]}^{1/2}, 
\label{phi1}
\end{equation}
where
\begin{equation} 
k=\Vert{\bf k}\Vert,\qquad q({\bf k})= N^{-1} \sum_j \phi_j e^{i {\bf k\cdot r}_j},
\label{phi2}
\end{equation}
${\bf r}_j$ is the position of site $j$, and ${\bf{k}} = (2\pi/L,0)$. 
Recall that this system is anisotropic, as interactions
along the spin axis are twice as large as in the perpendicular direction. Then, one 
could define a correlation length along the $Y$ axis, $\xi_{y,L}$, by choosing
${\bf{k}} = (0,2\pi/L)$. We have found that $\xi_{x,L}$ is more convenient because it is less affected by
finite--size effects than $\xi_{y,L}$. In order
to compare with similar quantities defined for isotropic systems like the short--range
2D EA Ising model, we define also $\xi_{L}=(\xi_{x,L}$+$\xi_{y,L})/2$.
In contrast to $P(q)$ and its first moments, $\xi_{L}$ takes into account spatial variations of the
EA overlap $q$ and shows a good signature of SG transition.
Its use has become customary in recent SG work.\cite{longi,3dEA}
Analogous expressions define the AF correlation length $\xi^{(m)}_L$ by substituting
$\phi_{j}$ for $\psi_{j}=(-1)^{x_i} \sigma_{i}$ in Eqs.(\ref{phi1}--\ref{phi2}).

It is worth mentioning that in the $\xi_L/L\to 0$ limit, $\xi_L$ is, up to a multiplicative
constant, the spatial correlation length of $\langle \phi_0\phi_r\rangle$.
Therefore in the paramagnetic phase we can think of $\xi_\infty$, the $L\to\infty$ limit of $\xi_L$, as the true
correlation length of a macroscopic system. On the contrary, if there is strong long--range  order with
short--range order fluctuations (as predicted for the droplet model \cite{droplet} for 3D SGs),
$q_2\neq 0$ (that is, $\langle \phi_0\phi_r\rangle$ does not vanish as $r\to \infty$) and
$\langle \phi_0\phi_r \rangle -\langle \phi_0\rangle  \langle\phi_r \rangle$ would be short--range.
It then follows from its definition Eq.(\ref{phi1}) that $\xi_L \sim L^2$ in 2D.
Following current usage, we shall nevertheless refer to $\xi_L$ as the ``correlation length'' .
\section{results}
\label{results}
\subsection{The AF phase}
\label{resultsA}

\begin{figure}[!t]
\includegraphics*[width=82mm]{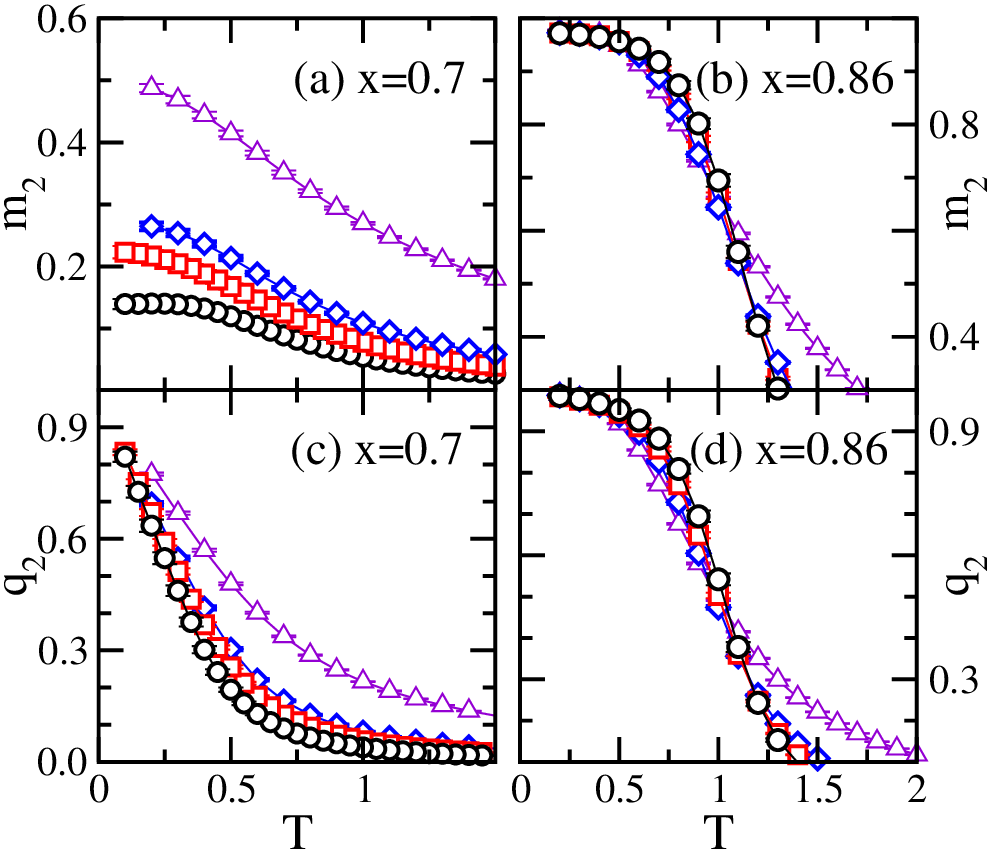}
\caption{(Color online) 
(a) Squared staggered magnetization $m_2$ vs $T$ for $x=0.7$. Icons $\circ$, $\square$, $\diamond$, and   $\triangle$ 
stand for $L=24,20,16$ and $8$ respectively. Lines are only guides to the eye.  Note that $m_2$ decreases with $L$
at all temperatures consistently with absence of  AF order. (b) Same as in (a) but for $x=0.86$.
Note that $m_2$ grows with $L$ at low temperature, indicating an AF phase. (c) Same as in (a)
but for the SG overlap parameter $q_2$. (d) Same as in (c) but for $x=0.86$. Direct comparison of curves
shown in panels (b) and (d) for $x=0.86$ indicate a coupling between $m_{2}$ and $q_{2}$.
This coupling does not occur for $x=0.7$ (see panels (a) and (c)).}
\label{magne}
\end{figure}

\begin{figure}[!b]
\includegraphics*[width=85mm]{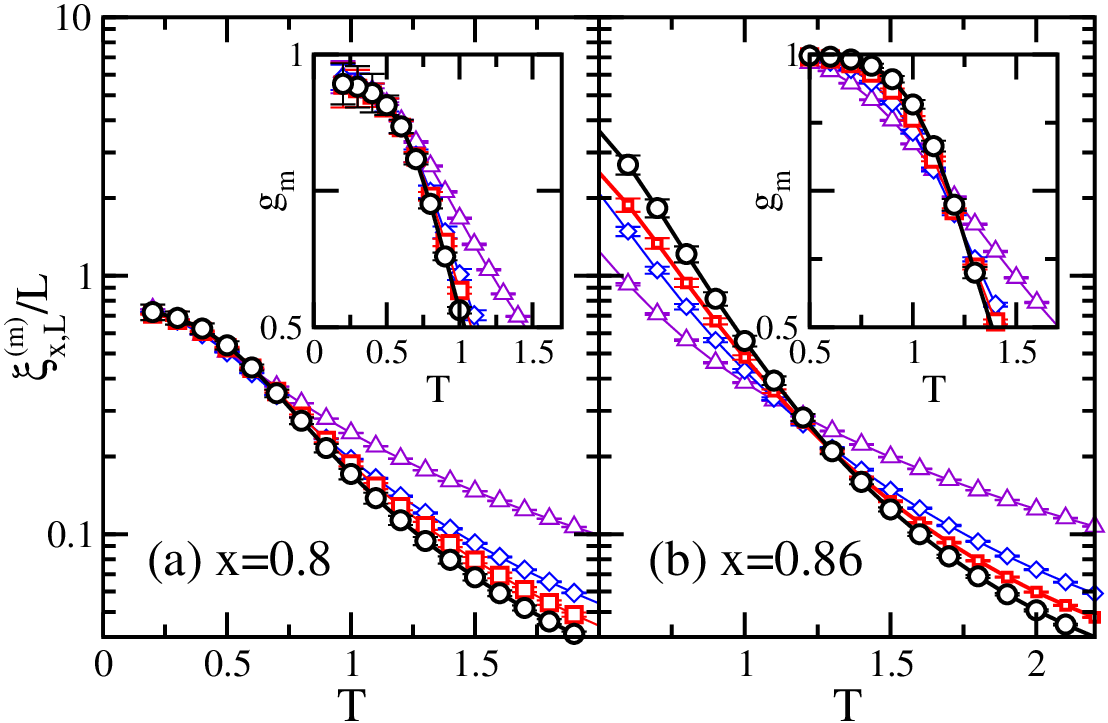}
\caption{(Color online)
(a) Semilog plots of $\xi^{(m)}_{x,L} /L$
versus $T$ for $x=0.8$, and 
$L=24$ ($\circ$), $L=20$ ($\square$), $L=16$ ($\diamond$),  and $L=8$ ($\triangle$). 
In the inset, kurtosis of the $m$ distribution versus $T$ for the same values of $x$ and
system sizes. (b) Same as in (a) but for $x=0.86$.  }
\label{longma}
\end{figure}

The phase diagram shown in 
Fig.~\ref{fases} summarizes our main results  for the diluted 2D PAD model.
For $x>x_{c}$ we find  a thermally  driven second order transition 
between the paramagnetic and AF phases at a N\'eel temperature 
$T_{N}(x)$ that vanishes as $x \to x_{c}$ from above. 
The phase boundary meets the $T=0$ line at $x_{c}\simeq 0.79$. For
concentrations well below $x_{c}$ the paramagnetic phase covers the
whole range $T \gtrsim  0$. We do not find evidence of a SG phase at finite
temperature. However, our results are consistent with a SG correlation
length that diverges algebraically near or at $T_{SG}=0$. In the following
we report the numerical evidence on which these qualitative results are based.

\begin{figure}[!b]
\includegraphics*[width=75mm]{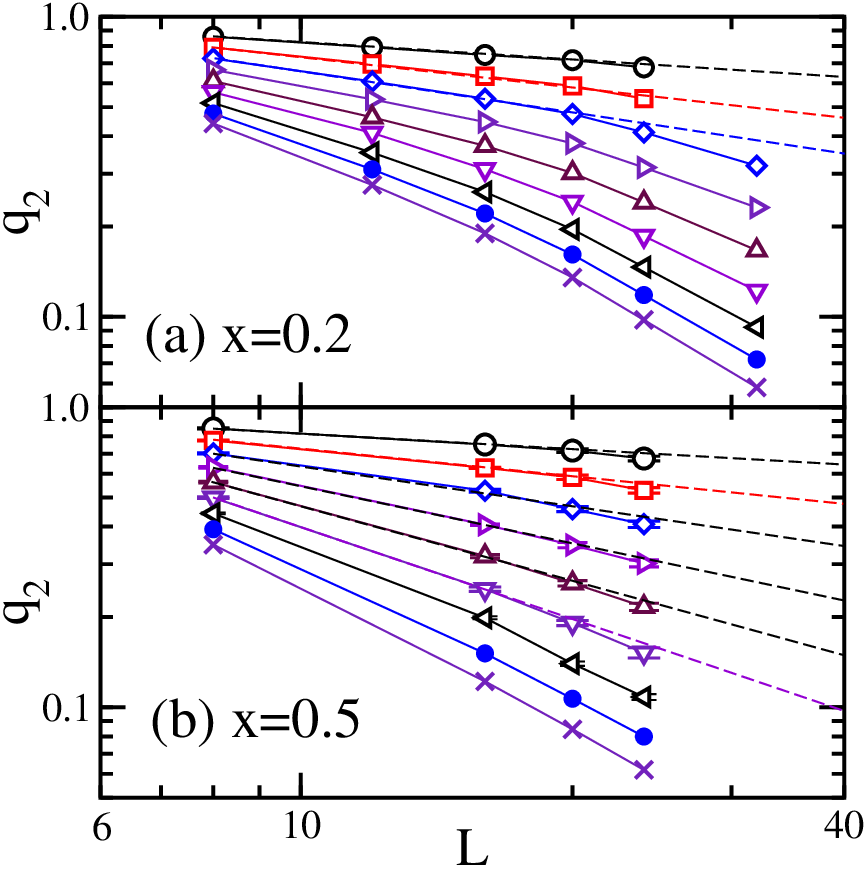}
\caption{ (Color online)
(a) Plots of $q_2$ versus $L$ for $x=0.2$. 
$\circ$, $\square$, $\diamond$, $\triangleright$, $\triangle$, $\triangledown$, $\triangleleft$, 
$\bullet$, and $\times$ stand for $T/x=0.2, 0.3, 0.4, 0.5, 0.6, 0.7, 0.8, 0.9$, and $1.0$
respectively. Lines are to guide the eye. Clearly, data for $T/x\gtrsim0.4$ deviate from
the straight dashed lines implying faster than a power of $1/L$ decay. (b) Same as in (a) but for $x=0.5$.
Here, only data for  $T/x\gtrsim0.7$ decay faster than a power of $1/L$.
For all data, we have checked that, within errors,  $\widetilde q_2=q_2$.}
\label{qus}
\end{figure}

First  we focus our attention on the paramagnetic--AF transition.\cite{DISx} The AF phase is
defined by the staggered magnetization~(\ref{phi}). We illustrate in Fig.~\ref{magne}a how 
the moment of staggered magnetization $m_2$ behaves with temperature for
$x=0.7$. Note that  $m_2$ appears to decrease as $L$ increases even at low
$T$. Plots of $m_2$ versus $L$ (not shown) indicate  a faster than
algebraic decay in $L$, as one expects for a non AF phase.
This is in sharp contrast to the behavior of $m_2$ for $x=0.86$ (see Fig.~\ref{magne}b).
Curves for different $L$ cross at $T_{N}\simeq1.15$. Below this temperature
$m_{2}$ increases with $L$ indicating the existence of an ordered AF phase.
Similar results are obtained for higher values of $x$. In the inset of Fig.~\ref{fases},
plots of $m_{1}$ versus $x$ for different system sizes at low
temperature show that the system exhibits AF order for $x  \gtrsim  0.8$.
Similar graphs were obtained for $m_{2}$.  These are our first pieces of evidence for 
the existence of an AF phase above $x_c\sim 0.8$. 
It is instructive to compare the behavior of $m_{2}$ with that of $q_{2}$ 
shown for $x=0.86$ in Figs.~\ref{magne}b and~\ref{magne}d.  $m_{2}$ and $q_{2}$
are not qualitatively different. This is not so for $x=0.7$ where there is no AF order
(compare Figs.~\ref{magne}a and~\ref{magne}c). From Fig.~\ref{magne}c
it is not obvious whether $q_{2}$  vanishes or not as $L$ increases at very low temperatures.
We will return to this point in the discussion of Fig.~\ref{qus}.

For further information about the extent of the AF phase, we also examine how
the cumulant--like quantity $g_{m}$ and the finite--size AF correlation length
behave for several pairs of values $x$, $T$. Let us first outline how $g_m$
is expected to behave in the various magnetic phases.
It clearly follows from its definition that $g_{m} \rightarrow 1$ as $L\rightarrow \infty$
in the case of long--range AF order. From the law of large numbers it also follows that
$g_{m} \rightarrow 0$ as $L\rightarrow \infty$ in the paramagnetic phase.
These two statements imply that curves of $g_m$ vs $T$ for
various values of $L$ cross at the phase boundary between the paramagnetic and AF
phases. We make use of this fact to quantitatively determine the paramagnetic--AF phase boundary.
Plots of $g_{m}$ vs $T$ are shown
in the insets of Figs.~\ref{longma}a and \ref{longma}b for $x=0.8$ and $0.86$, respectively.
The signature of an AF phase below $T \simeq 1.2$ is clear for $x=0.86$.
The inset of Fig.~\ref{longma}a shows that within errors curves
of $g_{m}$ vs $T$ for $x=0.8$  and different system sizes merge instead of crossing
at and below $T=0.5(1)$. Note also that $g_{m}$ does not go to $1$ as $T \to0$,
indicating a broad distribution of $P(m)$ even in this limit. Finally,
for $x<x_{c}$ (see Section~\ref{resultsB}), we find that $g_{m}$ decreases as $L$
increases  for all $T$ which is consistent with the absence of AF in this region.

\begin{figure}[!b]
\includegraphics*[width=85mm]{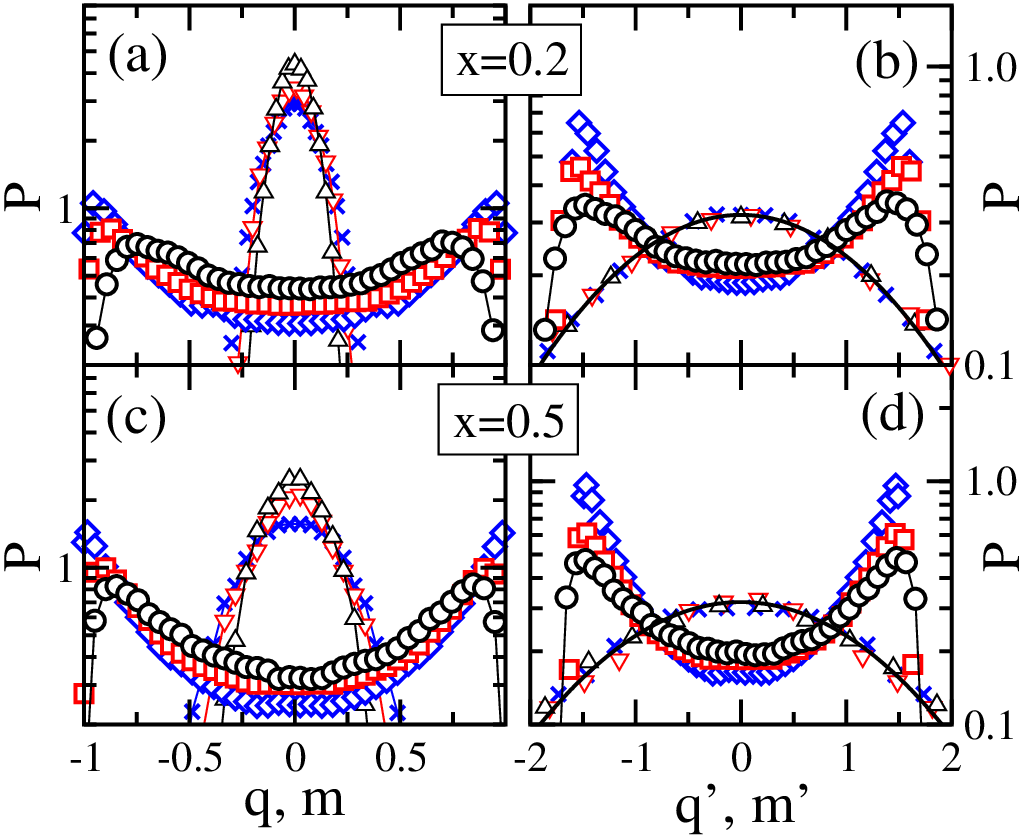}
\caption{  (Color online)
(a) Plots of the probability distributions $P(q)$ versus $q$, and $P(m)$ versus $m$ 
both for $x=0.2$ and $T/x=0.4$. 
$\circ$, $\square$, $\diamond$  are for $P(q)$  and system sizes $L=32, 24$ and 
$20$, respectively.  $\triangle$, $\triangledown$, and $\times$ are
for $P(m)$  and system sizes $L=24, 20$ and  $16$, respectively.
(b) Same as in (a) but for the scaled distributions $P(q^\prime)$ versus $q^\prime=q/q_{1}$,
and $P(m^\prime)$ versus $m^\prime=m/m_{1}$. The thick dashed line corresponds to the Gaussian 
distribution of paramagnets in the macroscopic limit.
(c) Same as in (a) but for $x=0.5$ and $T/x=0.4$. 
$\circ$, $\square$, $\diamond$  are for $P(q)$  and system sizes $L=24, 20$ and 
$16$, respectively.  $\triangle$, $\triangledown$, and $\times$ are for 
$P(q)$  and system sizes $L=24, 20$ and  $16$, respectively.
(d) Same as in (b) but for $x=0.5$ and $T/x=0.4$.
}
\label{histos}
\end{figure}

In recent literature on SG phases the scale invariant finite--size correlation length\cite{longi}
is frequently used to give evidence for a finite temperature transition, since  $\xi_L/L$ crosses at the
transition temperature $T_N$ and spreads out above and below $T_N$. 
The advantange of $\xi_L/L$ over kurtosis is that the former
may even diverge as $L  \to \infty$ in contrast to the latter that tends to~1.
Then we use the AF correlation length  $\xi^{(m)}_L/L$ to pinpoint values for $T_N$
by the value of $T$ where curves cross. Recall that $\xi^{(m)}_{L}$ becomes a true correlation
length when $\xi^{(m)} _{L}/L\ll 1$. Then, in the paramagnetic phase, $\xi^{(m)}_L/L\sim O(1/L)$,
therefore decreasing as $L$ increases. At $T_N$, $\xi^{(m)}_L/L$ must become size independent, as expected
for a scale--free quantity. At lower temperatures, well in the long--range AF phase, we
expect  $\xi^{(m)}_L/L\sim O(L)$. Plots of $\xi^{(m)}_{x,L}/L$ versus  $T$ are shown in
Figs.~\ref{longma}a and \ref{longma}b for  $x=0.8$ and $0.86$, respectively.
Note that curves spread out above and below $T_N =  1.20(5)$ for $x=0.86$. Similar
graphs for $x=0.9$ allow one to obtain the value $T_N = 1.50(5)$.

On the other hand curves 
merge for all temperatures below $T=0.5(1)$
for $x=0.8$ and  $L\ge16$ (see Fig.~\ref{longma}a), 
while $m_{2}$ decreases, within errors, algebraically
with $L$ for the studied system sizes (not shown). It is interesting to note that graphs of
the SG quantities $g$ and $\xi_L/L$ (instead of $g_m$ and the AF $\xi^{(m)}_L/L$)
give qualitatively the same picture when plotted versus $T$ except from
the fact that $g \to 1$ as $T \to 0$ for all $x$.
Thus, the most straightforward interpretation of the data shown in Fig.~\ref{longma}a
is that for all temperatures below $T=0.5$ the system is
near or at the AF phase boundary and its behavior displays criticality.

We have thus established all points of the AF phase boundary
shown in Fig.~\ref{fases} for $x> 0.7$. A fit to these data points, given by
$T_N \simeq 4.5 (x-x_c)^{1/2}$, where $x_c=0.79(5)$ is shown in Fig.~\ref{fases}.
Finally, for $x\le0.7$ (see below) we find that $\xi^{(m)}_L/L$ decreases as
$L$ increases for all $T$, as expected.
  
\subsection{Very diluted systems}
\label{resultsB}
This Section is devoted to show the numerical results drawn for distributions of $q$ and
their first moments, and for $\xi_L$ for systems with weak concentration.
As for 3D PADs,\cite{2009b} we expect universal behavior for
$x\ll 1$ which enables us to compare our results with previous work.
Thus, we direct our attention on the results we have obtained for
$x=0.2$ and $x=0.5$.  Both values are well below the $x \ge x_{c}=0.79$ region,
in which AF appears at low temperatures.

A plot of $q_2$ versus $T$ for $x=0.7$ is shown in Fig.~\ref{magne}c.
Qualitatively similar graphs are obtained for other values of $x$
satisfying  $x\lesssim x_c$. Note that  $q_2$ decreases
as $L$ increases, even at low temperatures. It is
difficult to deduce from these plots whether or not $q_2$ vanishes
as $L\rightarrow \infty$ at low $T$. In order to elucidate this question
we prepare log--log plots of $q_2$ vs $L$, shown in  Figs.~\ref{qus}a and~\ref{qus}b
for $x=0.2$ and $0.5$ respectively.
Data points in these figures seem to be consistent with a decay faster than
$q_2\sim L^{-\eta}$ for $T/x \gtrsim 0.3$, indicating that we are in the paramagnetic phase.
Plots of  $q_1$  vs $L$ show the same qualitative behavior.
Altogether these results leave small room for the existence of a SG phase with
quasi--long range order at very low temperatures, as it has been reported for
the 3D diluted PAD model for $T/x \gtrsim 1$.

\begin{figure}[!t]
\includegraphics*[width=70mm]{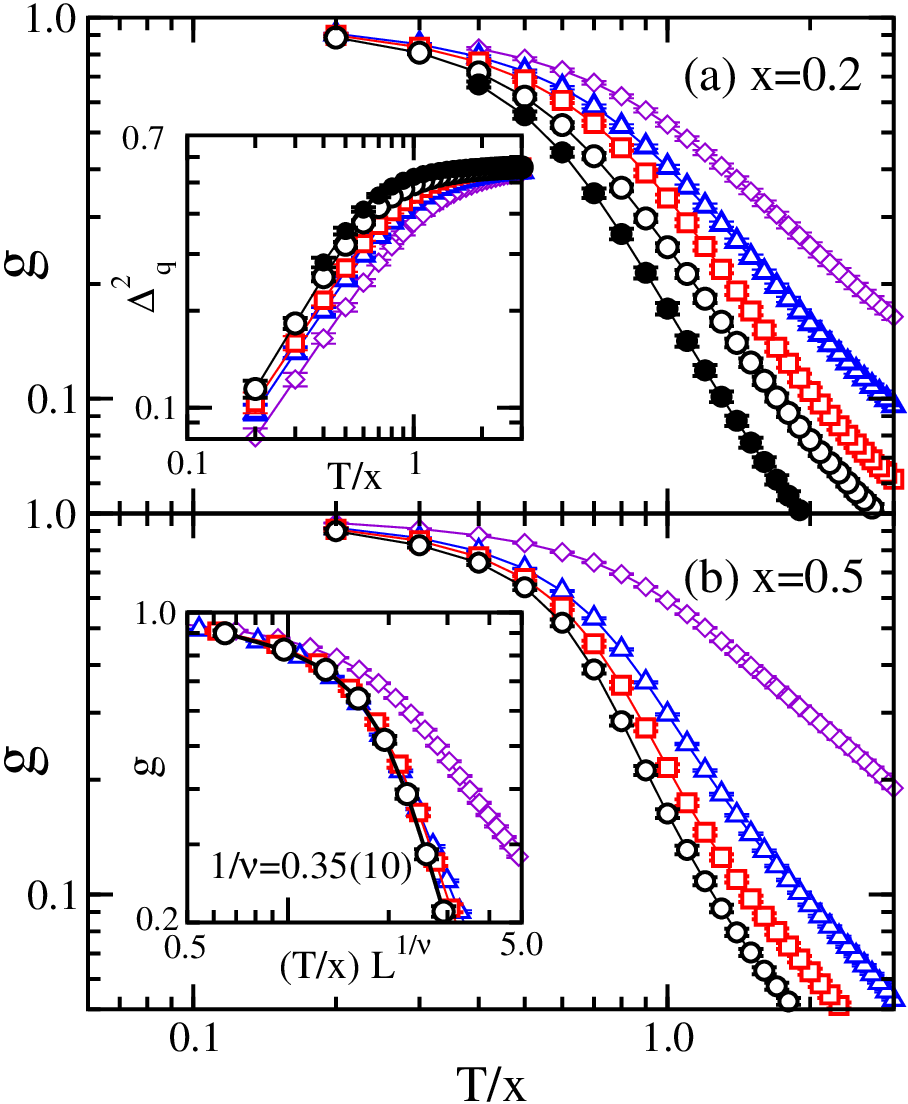}
\caption{ (Color online)
Kurtosis $g$ of the distribution of $q$ versus $T/x$ for $x=0.2$. 
$\bullet$, $\circ$, $\square$, $\triangle$, and $\diamond$ stand
for system sizes $L=32, 24, 20,16$, and $8$ respectively. 
In the inset, semilog plot of the relative mean square deviation 
$\Delta_q^2$ versus $T/x$ for the same $x$ and system sizes.
Lines are guides to the eye. 
(b) Same as in (a) but for $x=0.5$.
$\circ$, $\square$, $\triangle$, and $\diamond$,  
stand for $L=24,20,16$, and $8$ respectively. In the inset, scaling plot 
($g$ as a function of $(T/x)L^{1/\nu}$) of the data shown in the main
figure.} 
\label{kurt}
\end{figure}

\begin{figure}[!t]
\includegraphics*[width=75mm]{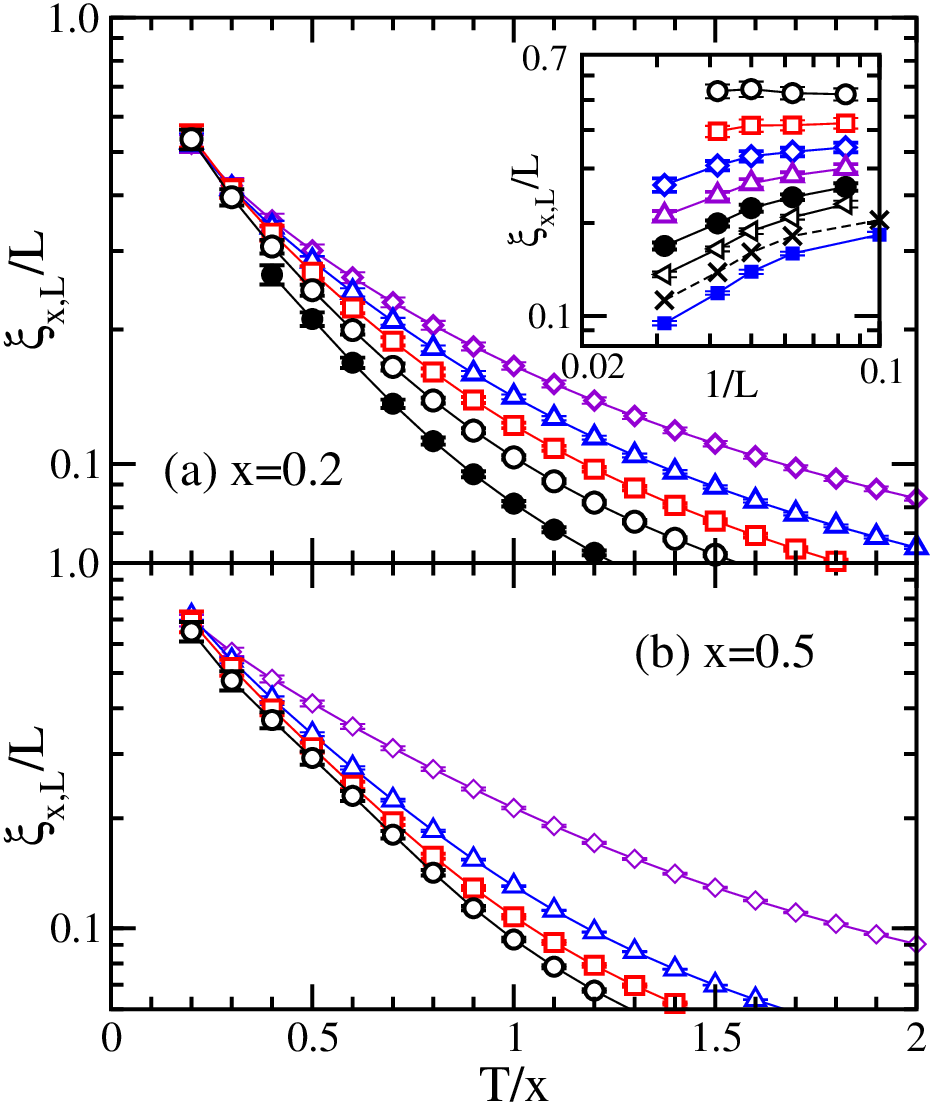}
\caption{(Color online)
(a) Semilog plots of (a) SG correlation length divided by system size $\xi_{x,L}/L$
versus $T/x$ for $x=0.2$, and 
$L=32$ ($\bullet$), $L=24$ ($\circ$), $L=20$ ($\square$), $L=16$ ($\triangle$),  
$L=12$ ($\diamond$), and $L=8$ ($\times$). Lines are guides to the eye.
(b) Same as in (a) but for $x=0.5$, and $L=24$ ($\circ$), $L=20$ ($\square$), 
$L=16$ ($\triangle$), and $L=8$ ($\diamond$). In the inset of Fig.~(a),
log--log plot of $\xi_{x,L}/L$ versus $1/L$ for $x=0.2$. $\circ$, $\square$,
$\diamond$, $\triangle$, $\bullet$, $\triangleleft$, $\times$, and $\blacksquare$ stand for 
$T/x=0.2, 0.3, 0.4, 0.5, 0.6, 0.7, 0.8$, and $1.0$ respectively.
}
\label{longqus}
\end{figure}

Next we report results for the distributions of $m$ and $q$ at low temperature.
Due to the central limit theorem and since the correlation lengths are finite in the
paramagnetic phase, $P(m)$ and $P(q)$ are expected to be normal distributions for $L=\infty$.
The droplet picture \cite{droplet} for SG's predicts that $P(q)=[\delta(q+q_0)+\delta(q-q_0)]/2$
where $q_0$ is the EA order parameter, and that the tail
of $P(q)$ down to $q=0$  for finite--size systems vanishes as $L$ increases. On the contrary, the
RSB picture \cite{RSB} predicts a nontrivial distribution with a non--vanishing  $P(q=0)$
which is size independent.  Plots of $P(q)$ and $P(m)$ are shown for $x=0.2$ and $x=0.5$
for $T/x=0.4$ in Figs.~\ref{histos}a--\ref{histos}c.
All distributions depend on $L$. $P(m)$ are found to be normally distributed.
In Figs.~\ref{histos}b and \ref{histos}d normalized distributions of the reduced
quantity $m^\prime=m/m_1$ are shown. Note that $P(m^\prime) \simeq (1/\pi) \exp(-m^{\prime\,2}/\pi)$
for the studied system sizes, indicating complete absence of AF order.
On the other hand, $P(q)$ are found to be double peaked distributions. As $L$ increases, their peak
positions shift towards $q=0$ and $P(0)$ increases.
Neither the droplet nor the RSB models consent a fit to these data.
If it turns out that our systems are near or at criticality, then $P(q^\prime)$ (where
$q^\prime=q/q_1$) ought to be size independent. However, reduced distributions
$P(q^\prime)$ in Figs.~\ref{histos}b--\ref{histos}d
are shown to have an $L$ dependence. We conclude that our results are only consistent
with a paramagnetic phase. Similar conclusions apply for  $T/x\gtrsim0.2$.

In the same way as explained in Section~\ref{resultsA} for quantities $g_{m}$
and $\xi^{(m)}_{L}/L$, their dimensionless SG counterparts $g$ and $\xi_{L}/L$, as well as $\Delta_q^2$,
indicate the location of the temperature $T_{SG}$ of a SG transition.
Recall that, according to the finite--size scaling assumption, all these quantities depend only on
$L(T-T_{SG})^\nu$ and then become size independent at $T_{SG}$. We are assuming that
for large enough sizes, $L/\xi$ (where  $\xi$ is the {\it true} correlation length) is the only relevant
parameter and $\xi \sim (T-T_{SG})^{-\nu}$. Plots of $g$ vs $T/x$ are given in Fig.~\ref{kurt}a and
\ref{kurt}b for $x=0.2$ and $0.5$ respectively. It seems that curves of $g$ for various values of $L$
do not cross and only merge as  $T \to 0$. This is consistent with $T_{SG}=0$ in accordance
with the behavior of 2D EA systems, although a merging at $T/x\approx 0.2$ is not completely excluded within errors.
We found more useful to study $\Delta_q^2$,  which has a direct interpretation as the uncertainty of $q/q_{1}$ and
could be computed with  higher precision as it involves lower moments of\cite{rad} $P(q)$. A plot
of $\Delta_q^2$ vs $T/x$ for $x=0.2$ is shown in the inset of Fig.~\ref{kurt}a. Clearly,
$\Delta_q^2$ increases with $L$ for all $T/x\ge0.2$ as expected for a paramagnetic
phase and contrary to the expected behavior $\Delta_q^2 \to 0$ for increasing $L$ in
a SG phase with non--vanishing order parameter. A similar behavior for $x=0.5$ is observed (not shown).

We now turn our interest to the behavior of the SG finite--size correlation length $\xi_{L}/L$. Similarly to $g$, $\xi_{L}/L$
is independent of $L$ at $T_{SG}$ as it corresponds to be for a scale--free quantity.  On the other hand, 
well inside the paramagnetic phase, $\xi_{L}/L$ should diminish as $O(1/L)$, provided $\xi_{L}/L\ll1$.
Figs.~\ref{longqus}a and~\ref{longqus}b show data for
$\xi_{x,L}$ as a function of $T/x$ for several system sizes, for $x=0.2$ and $0.5$ respectively. 
Curves do not cross at any finite temperature and
$\xi_{x,L}/L$ decrease as $L$ increases at least for $T/x \gtrsim 0.4$.
All curves seem to converge only as $T \to 0$ suggesting $T_{SG}=0$.
Plots for $\xi_{x,L}$ vs $L$ (see inset of Fig.~\ref{longqus}a) are consistent with an algebraic decay 
$\xi_{x,L}/L \sim L^{-y}$ for temperatures $T/x \gtrsim 0.4$ and system sizes $L\ge16$. For
lower temperatures, data do not vary very much as $L$ increases and so we cannot definitely rule 
out a non--vanishing $\xi_{x,L}/L$ in the thermodynamic limit. Consequently, from Fig. \ref{longqus} alone,
a transition at very low, but not zero, temperature cannot be completely excluded.

\begin{figure}[!t]
\includegraphics*[width=80mm]{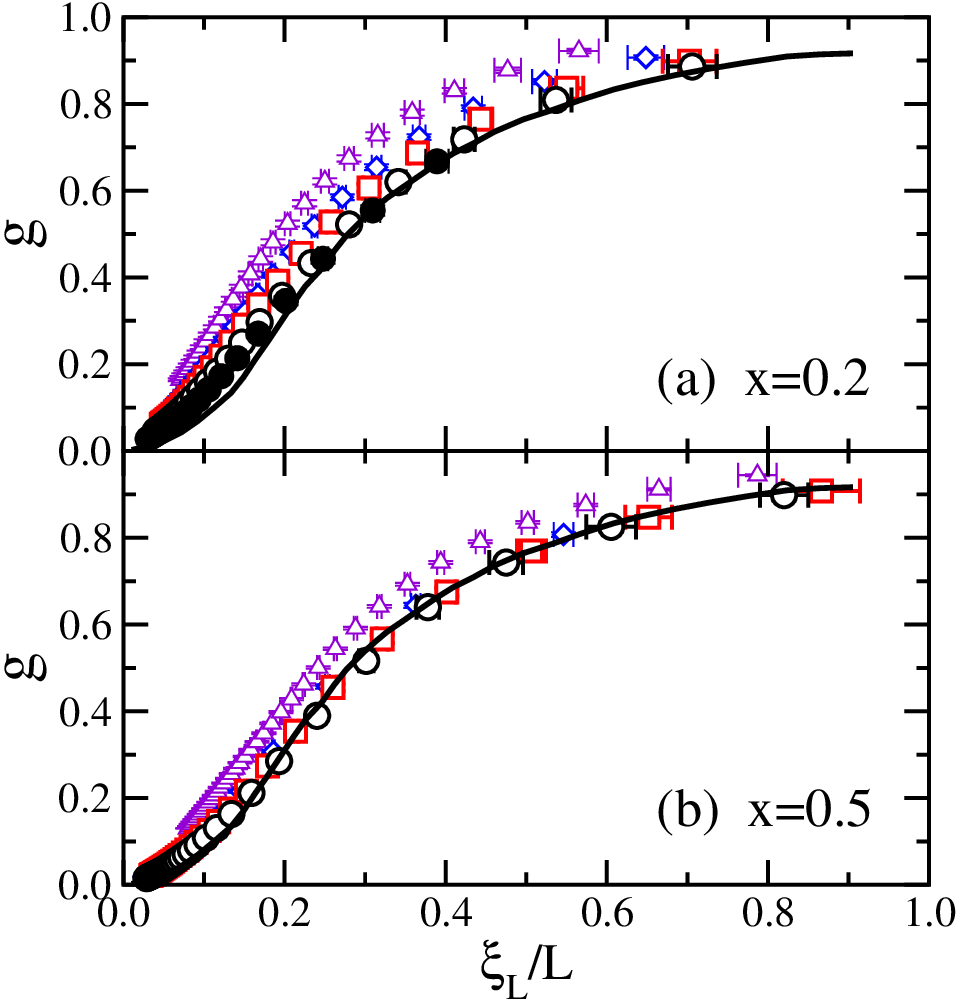}
\caption{(Color online)
(a) Kurtosis $g$ as a function of 
the finite--size correlation length divided by system size, 
$\xi_{L}/L$ for $x=0.2$. $\bullet$, $\circ$, $\square$, $\diamond$, and $\triangle$ stand
for $L=32, 24, 20,16$, and $12$ respectively. All data should collapse onto an universal curve,
provided that scaling corrections are small. The thick continuous line stands for
the universal curve that corresponds to the 2D Ising SG model with short--range interactions.
(b) Same as in (a) but for $x=0.5$. $\circ$, $\square$, $\diamond$, and $\triangle$ stand
for system sizes $L=24, 20,16$, and $8$ respectively.
}
\label{katz}
\end{figure}

\begin{figure}[!t]
\includegraphics*[width=80mm]{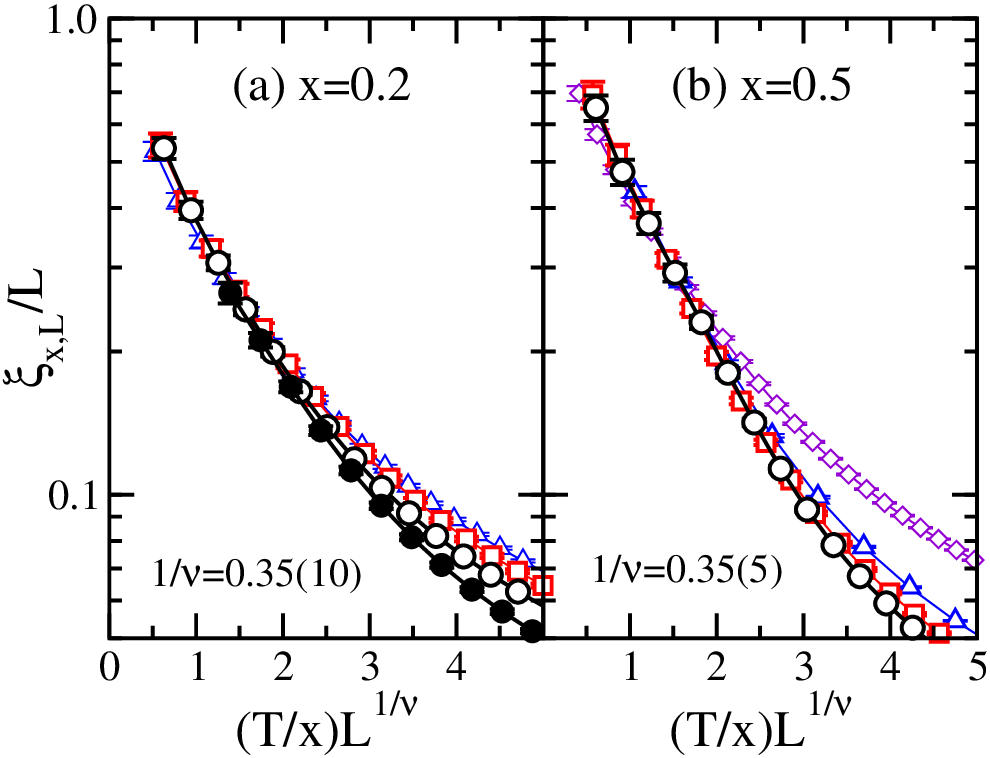}
\caption{(Color online) (a) Scaling plot for the SG correlation length divided by system size,
$\xi_{x,L}/L$ versus $(T/x) L^{1/\nu}$ with $1/\nu=0.35$ for $x=0.2$. 
$\bullet$, $\circ$, $\square$, and $\triangle$ are for $L=32, 24, 20$ and $16$, respectively.
(b) Same as in (a) but for $x=0.5$.  
$\circ$, $\square$, $\triangle$,  and $\diamond$ are for $L=24, 20, 16$ and $8$, respectively.
Error bars, where not shown, are smaller than symbols.}
\label{longescala}
\end{figure}

\subsection{The $\nu$ exponent at $T_{SG}=0$}
\label{resultsC}

Our results concur with the behavior found for a random--bond 2D RAD model with dipolar
interactions,\cite{rad} and the one found  time ago for 2D EA models.\cite{2dEAold} Recent
simulations for the latter (with larger system sizes and lower $T$) including nearest--neighbor
exchange interactions find a lower value $1/ \nu = 0.29(4)$ for Gaussian
distributions,\cite{2dEAnew} but do not provide conclusive results for bimodally distributed
interactions.\cite{bis} A scenario has been proposed where these models for
varying realizations of disorder belong to the same universality class at non--zero temperatures,\cite{jorg}
even though their respective behaviors differ at $T=0$.

Let us assume that $T_{SG}=0$ and that $\xi$ diverges as $\xi \sim T^{-\nu}$. 
According to finite--size scaling, dimensionless quantities like $g$ and $\xi_{L}/L$
should scale as

\begin{equation} 
g=G(T L^{1/\nu}), ~~\xi_{L}/L=X(T L^{1/\nu}).
\label{q2}
\end{equation}
It follows that $g=F(\xi_{L}/L)$, where $F$ is in principle a non--universal function that,
apart from the bulk universality class, depends on the boundary conditions (we chose them to be
periodic), the sample shapes (squared lattices), as well as the anisotropy\cite{shchur} of
the interactions. Plots of $g$ versus $\xi_{L}/L$ are shown in
Figs.~\ref{katz}a and~\ref{katz}b for $x=0.2$ and $0.5$ respectively.
Data for different values of $L$ and $x$ should collapse into a single
scaling curve, on condition that finite--size effects are small. This is what we find
for systems with $N \gtrsim 100$. However, for smaller $N$ curves spread out
indicating that finite--size scaling corrections are large.
It is remarkable that data seems to collapse onto the universal scaling curve
found for the (isotropic) 2D EA models mentioned above in this Section,\cite{2dEAnew,bis}
specially at low $T$.  Note that in Figs.~\ref{katz}a and~\ref{katz}b we use $\xi_{L}/L$
instead of $\xi_{x,L}/L$ in order to average  anisotropic effects over the two principal axes 
of the underlying square lattice.
This suggests that both, 2D PAD and short--range EA models, may share a common
universality class. However, large correlations to scaling for the available system sizes
prevent us to  go further on this direction. 

Scaling plots of  $\xi_{x,L}/L$ versus $(T/x) L^{1/\nu}$ are shown in
Figs.~\ref{longescala}a and~\ref{longescala}b for $x=0.2$ and $x=0.5$ respectively.
Due to the presence of large finite--size corrections, no value of $1/\nu$ allows 
to collapse all data in one single curve.
We have chosen $1/\nu$ in order to allow data collapse for large $L$ and low
$T$. We find that $1/\nu=0.35$ is a suitable value in both cases.
Scaling plots become significantly worse (not shown) when using values of $1/\nu$ 
outside the interval  $[0.25,0.45]$ ($[0.3,0.4]$) for $x=0.2$ ($x=0.5$). 
That gives a rough estimate of the error on $1/\nu$. A scaling plot of 
$g$ versus $(T/x) L^{1/\nu}$ is shown in the inset of Fig.~\ref{kurt}b for $x=0.5$. Again, 
$1/\nu=0.35(10)$ allows to scale data for large $L$ and low $T$, which is consistent 
with the value extracted from
$\xi_{x,L}/L$. The value $1/\nu=0.35(10)$ also agrees with the effective exponent found in
old simulations of the 2D EA model for small system sizes and relatively high
temperatures and is slightly larger than the value $1/\nu=0.29(4)$ found in recent simulations
for the same models (but the two values are still consistent within errors).
In any case, such discrepancies may be caused by the fact that finite--size scaling corrections are
important for the limited system sizes that we were able to simulate.

\section{conclusions}
\label{conclusions}

By tempered Monte Carlo calculations, we have studied a diluted dipolar 
Ising model on a square lattice.  There are only dipole--dipole interactions.
Spins randomly occupy only a fraction $x$ of all lattice sites.
The entire phase diagram of the system, in particular the boundary between the
AF and the paramagnetic phases, has been explored and it is shown in Fig.~\ref{fases}.
We have also provided strong evidence that the paramagnetic phase covers the whole
$T > 0$ range for $x < x_c$, where $x_c= 0.79(5)$.
From the behavior of  the spin--glass (SG) overlap $q$, the relative mean square 
deviation $\Delta_q^2$, kurtosis $g$, and $\xi_L/L$, we conclude that $T_{SG}=0$ for 
$x< x_c$, and there is an algebraic divergence of the correlation length with an
exponent $1/\nu=0.35(10)$. All these properties are consistent with the behavior found for the
2D diluted RAD and 2D EA model with short--range interactions. This is to be contrasted with the
manifestly different behavior found for the 3D PAD (quasi long--range order at low $T$) and
the 3D EA model (with a non--vanishing order parameter in the SG phase).

\acknowledgments
We thank Julio F. Fern\'andez for helpful discussions and reading the manuscript.
We are indebted to the Centro de Supercomputaci\'on y Bioinform\'atica and to the
Applied Mathematics Department both at University of M\'alaga (Spain), to
Institute Carlos I at University of Granada (Spain) and to the SP6 computer array at
CINECA in Bolonia (Italy) for much computer time. JJA thanks financial support from
a Junta de Andaluc\'{\i}a Grant FQM--278--2010.

\end{document}